\documentclass[manuscript]{aastex}
\usepackage[utf8]{inputenc}
\usepackage{graphicx}
\usepackage{amsmath} 
\usepackage{wasysym} 
\usepackage{bm}  
\usepackage{ulem} 

\newcommand{\bmu}{{\bm{\mu}}}
\newcommand{\bpi}{{\bm{\pi}}}
\newcommand{\bgamma}{{\bm{\gamma}}}

\newcommand{\up}[1]{\ifmmode^{\rm #1}\else$^{\rm #1}$\fi}

\newcommand{\arcd}{\ifmmode^{\circ}\else$^{\circ}$\fi}
\newcommand{\arcm}{\ifmmode{'}\else$'$\fi}
\newcommand{\arcs}{\ifmmode{''}\else$''$\fi}

\slugcomment{\today{} version}

\shorttitle{OGLE-2012-BLG-0406Lb}
\shortauthors{Poleski et al.}

\begin{document}


\title{Super-massive planets around late-type stars -- the case of OGLE-2012-BLG-0406Lb}

\author{Radosław Poleski\altaffilmark{1,2}, Andrzej Udalski\altaffilmark{2}, Subo Dong\altaffilmark{3},}
\author{Michał K. Szyma\'nski\altaffilmark{2}, Igor Soszyński\altaffilmark{2}, Marcin Kubiak\altaffilmark{2}, Grzegorz Pietrzyński\altaffilmark{2,4},}
\author{Szymon Kozłowski\altaffilmark{2}, Paweł Pietrukowicz\altaffilmark{2}, Krzysztof Ulaczyk\altaffilmark{2}, Jan Skowron\altaffilmark{2},}
\author{Łukasz Wyrzykowski\altaffilmark{2,5} and 
Andrew Gould\altaffilmark{1}}
\email{poleski@astronomy.ohio-state.edu}

\altaffiltext{1}{Department of Astronomy, Ohio State University, 140 W. 18th Ave., Columbus, OH 43210, USA}
\altaffiltext{2}{Warsaw University Observatory, Al. Ujazdowskie 4, 00-478 Warszawa, Poland}
\altaffiltext{3}{Institute for Advanced Study, Einstein Drive, Princeton, NJ 08540, USA}
\altaffiltext{4}{Universidad de Concepci\'on, Departamento de Astronomia, Casilla 160–C, Concepci\'on, Chile}
\altaffiltext{5}{Institute of Astronomy, University of Cambridge, Madingley Road, Cambridge CB3 0HA, UK}


\begin{abstract}
The core accretion theory of planetary formation does not predict that super-Jupiters will form beyond the snow line of a low mass stars. 
We present a discovery of $3.9\pm1.2~M_{\rm Jup}$ mass planet orbiting the $0.59\pm0.17~M_{\odot}$ star using the gravitational microlensing method. 
During the event, the projected separation of the planet and the star is $3.9\pm1.0~{\rm AU}$ i.e., the planet is significantly further from the host star than the snow line. 
This is a  
fourth 
such planet discovered using the microlensing technique and challenges the core accretion theory. 
\end{abstract}



\section{Introduction} 

There are two main theories of planetary formation: core accretion and gravitational instability.
The first one does not predict planets to be formed much beyond the snow line 
(the ring in protoplanetary disks where the temperature is below the sublimation temperature of ice). 
One of the tests for these theories is the observational census of the super-Jupiter mass planets around low-mass stars, i.e., M dwarfs \citep{laughlin04}. 
\cite{kennedy08} predicted that the probability that an $0.6~{M_\odot}$ star has at least one giant planet is $2\%$. 
The slope $\alpha$ of the planetary mass function ($dN/d\ln M\sim M^\alpha$) is $-0.31\pm0.2$ 
according to \citet{cumming08}, for other estimates, see, e.g., \citet{cassan12}. 
For super-Jupiters, the planetary mass function should be even steeper because their mass is close to the total mass of the disk. 
We may expect that the probability that an $0.6~{M_\odot}$ star has a $2-4~M_{\rm Jup}$ mass object is $\approx1\%$ or even smaller.
Thus, detection of a super-Jupiters beyond the snow line of a late-type stars challenges the core-accretion theory.

It is hard to discover super-Jupiters around low-mass stars 
using either radial velocities or transit methods if the planetary orbit is beyond the snow line.
For a $0.5M_{\odot}$ star, the snow line is located at about $1.3~{\rm AU}$. 
If the planet is twice further, then its orbital period is $5.9~{\rm yr}$ 
and the radial velocity amplitude is $25(M_{p}/M_{\rm Jup})~{\rm m~s^{-1}}$, 
where $M_p$ is the mass of the planet.
Although the amplitude is large, the long orbital period of the planet makes detection difficult. 
None of the radial velocity detected planets is a super-Jupiter beyond the snow line on 
a list of known planets around M dwarfs presented by \citet{bonfils13}. 
Though, close to this part of the 
parameter 
space there are planets GJ 317b and GJ 676Ab. 
Among the planetary transit candidates with more than one transit announced by the {\it Kepler} mission,  
the longest orbital periods are around $500~{\rm d}$ \citep{borucki11,batalha13}, i.e., much shorter than the expected $5.9~{\rm yr}$. 
Moreover, the probability of the proper orbit alignment is very low; 
hence, it is very difficult for transit surveys to find planets with $5.9~{\rm yr}$ periods if they exist.  
Also high spatial resolution imaging is not an efficient way of finding such planets, as they are too close to the parent star. 
Thus, gravitational microlensing seems to be the most efficient method in discovering super-Jupiter 
planets around dwarf stars \citep{gaudi12}. 
The relatively large mass ratio $q$ makes both central and planetary caustics large, 
as their size is proportional to $q$ and $q^{1/2}$, 
respectively \citep{gould92,griest98,chung05,han06}. 
This makes discovering such planets more feasible, because the probability that the background source would pass close enough to the caustic is higher. 

In fact, microlensing has already been successful in discovering massive planets around M dwarfs. 
\citet{dong09a} analyzed all available data, including the Hubble Space Telescope images, to constrain the parameters of OGLE-2005-BLG-071Lb, which was discovered by \citet{udalski05}.
\citet{dong09a} concluded that the $3.8 \pm 0.4~M_{\rm Jup}$ mass planet orbits an $M = 0.46\pm0.04~M_{\odot}$ star and lies at projected separation of $3.6\pm0.2~{\rm AU}$.
\citet{batista11} analyzed the MOA-2009-BLG-387 event. 
The projected separation of the planet and the star was comparable to the size 
of the Einstein ring, which resulted in a very large resonant caustic.
This allowed a very precise measurement of the mass ratio $q = 0.0132 \pm 0.003$ with most probable planet and host masses of $2.6~M_{\rm Jup}$ and $0.19~M_{\odot}$, respectively.
The estimated projected separation is $1.8~{\rm AU}$. 
There are two more microlensing planets that were claimed to be super-Jupiters orbiting low-mass stars beyond the snow-line based on Bayesian analysis, not the direct measurement of lens mass from microlensing model: OGLE-2003-BLG-235Lb \citep{bond04,bennett06} and MOA-2011-BLG-293Lb \citep{yee12}. 
Both events were observed with high spatial resolution, and these observations neglected the results of Bayesian analysis in the latter case \citep{batista13}.
We note that \citet{street12} found a planetary mass object 
MOA-2010-BLG-073Lb
($11~M_{\rm Jup}$) orbiting an M-dwarf beyond the snow line, but they concluded that it is most likely extremely low-mass product of star formation. 

The observing strategy which resulted in almost all of the microlensing planets discovered so far involved survey and follow-up groups. 
Survey groups conduct observations with 1-m class telescopes and large CCD cameras.
Microlensing Observations in Astrophysics (MOA-II) is using 
an $80~{\rm Mpix}$ camera with $2.2~{\rm deg^2}$ field of view.
The camera used during the fourth phase of the Optical Gravitational Lensing Experiment (OGLE-IV) 
contains $256~{\rm Mpix}$ and gives  $1.4~{\rm deg^2}$ field of view.
The search for microlensing events is performed by survey groups on a daily basis, and all the events found are presented to the astronomical community.
Then, the follow-up groups, which have access to a large number of different class telescopes 
widely spread in geographic coordinates, observe the events that are likely to be highly magnified 
(thus more sensitive to planets) or already show anomalies deviating from the standard microlensing light curve. 
Until now only two secure planets were announced using survey-only data 
\citep[OGLE-2003-BLG-235Lb/MOA-2003-BLG-53, and MOA-2007-BLG-192Lb][respectively]{bond04,bennett08} 
and both of them used MOA and OGLE data. 
\citet{bennett12} analyzed the microlensing event MOA-bin-1 using only MOA data and concluded that the lensing star hosts a planet. 
However, this case lacks the strong lensing signal of the host star. 
Thus, the host lensing parameters and the fact that the event was caused by microlensing do 
not rely on the data other than the anomaly. 
We note that sources other than planets can cause anomalies in 
microlensing 
events, see, e.g., \citet{gould13a} for MOA-2010-BLG-523.
\citet{yee12} showed that the planet MOA-2011-BLG-293Lb would have been discovered in survey-only data, 
if photometry from OGLE-IV, MOA-II, and Wise Observatory were routinely combined and analyzed jointly for all microlensing events.
However, this still has not been implemented. 
The currently operating and future surveys have greatly increased the number of events discovered;  
hence, the follow-up groups are unable to cover all the ongoing events. 
Also, the cadence of survey observations increased.  
The consequence of these is an increase in the number of planetary detections in survey-only data. 
At this point, the questions about the reliability of detections and the accuracy of parameters found in survey-only data arises \citep{yee12}.

Here we present the detection of a planetary system in the microlensing event OGLE-2012-BLG-0406. 
We use the data collected by the OGLE survey only. 
The planet turned out to be a super-Jupiter orbiting a star that is an M or K dwarf. 
The projected separation places the planet well beyond the snow line of the host star,
making it the  
fourth
such object discovered via microlensing.

\section{Observations} 

The observations were collected using the 1.3-m Warsaw Telescope situated in Las Campanas Observatory, Chile. 
The observatory is operated by the Carnegie Institute for Science. 
The telescope is equipped with a 32 CCD chip mosaic camera, which gives $1.4~{\rm deg}^2$ field of view. 
The pixel size is $15~\mu{\rm m}$, corresponding to $0.26\arcs$ on the sky. 
The observations were conducted using the $I$-band filter. 
A few $V$-band observations of different fields are taken each night.
They do not constrain the microlensing model and are used only to assess the color of the source star. 

The microlensing event OGLE-2012-BLG-0406 was found at equatorial coordinates of 17:53:18.17 $-30$:28:16.2
($l = -0.46^{\circ}$, $b = -2.22^{\circ}$) 
by the Early Warning System \citep{udalski03}. 
The event was found in the OGLE-IV field that is 
monitored with a typical cadence of $55$ minutes. 
The photometry for nights at ${\rm HJD'} = {\rm HJD} - 2450000$ of 6108 and 6109 showed a significant disagreement 
with a previously fitted standard \citet{paczynski86} light-curve. 
Then the observations of the field were suspended because of proximity of the Moon. 
The microlensing community was alerted on ${\rm HJD'} = 6110.8099$ that the event
is showing an anomaly by V.~Bozza who browsed the OGLE Early Warning System alert webpage, 
which presents updated OGLE photometry. 
The alert prompted the follow-up groups to collect additional
measurements, which are not analyzed here.
It is worth noticing here, that
OGLE-IV microlensing strategy in the high cadence fields is to
conduct well defined automated experiment with minimum human
intervention; thus, the OGLE-IV by definition does not distribute anomaly alerts. 

The OGLE observations returned at ${\rm HJD'} = 6113.52784$ and showed the star brightened by $0.3~{\rm mag}$.
The following images revealed it was also fading. 
This made it obvious that the additional peak in the light curve occurred when the Moon was passing close to the event.
Further observations were carried out with the standard cadence except that one additional measurement was taken at ${\rm HJD'} = 6113.85679$, i.e., 25 minutes after the previous one, to secure additional data point
of the setting lens before reaching the OGLE high air mass limit.
This does not significantly affect the microlensing model fitted and thus the analysis 
can be treated as based on survey-only data. 
We note that observations of this event were also taken by follow-up groups during the Moon gap in the OGLE data,
and these can be used to verify our results. 
After the Moon gap, the anomaly lasted for ~10 days  including one
more peak centered on ${\rm HJD'}=6120.9$. 

The main peak of the event occurred
at 
$t_0 = 6141.5$.  
It was just after the Moon passed close to the OGLE-2012-BLG-0406. 
Unfortunately, a suspicious linear small scale
trend is present in our photometry 
during the four nights starting from ${\rm HJD'} = 6140$. 
Larger scatter during that time is also present in photometry of nearby red clump
stars and in some cases linear
trends with $2\sigma$ significance are seen. High sky background or poor seeing
conditions (above $1\farcs7$) clearly affected at least nine out of $21$ epochs
collected at that time (including all on ${\rm HJD'} = 6142$ and $6143$).
It is worth noticing, that $1\farcs5$ away from the source star there is another star that is $1.1~{\rm mag}$ brighter than the event in
the baseline and was $0.3~{\rm mag}$ brighter than the event during the main
peak. 
This neighbor star most likely affected the event photometry,
especially at poor seeing conditions. As these factors caused not only
larger scatter of points but also an apparent linear trend, we decided
to skip these data in our primary fitting procedure and check their
influence separately.
During $2t_{\rm E} = 128.7~{\rm d}$ period centered on $t_0$ altogether 528 $I$-band data points were collected 
($t_{\rm E}$ is Einstein crossing time, $t_0$ is time of the closest approach). 
The sky area, where the event is located, was also observed during the third phase of the OGLE project. 
Altogether, more than 4300 epochs were collected since 2001, and no other variability was seen except the 
analyzed event. 

\begin{figure}
\epsscale{.764}
\plotone{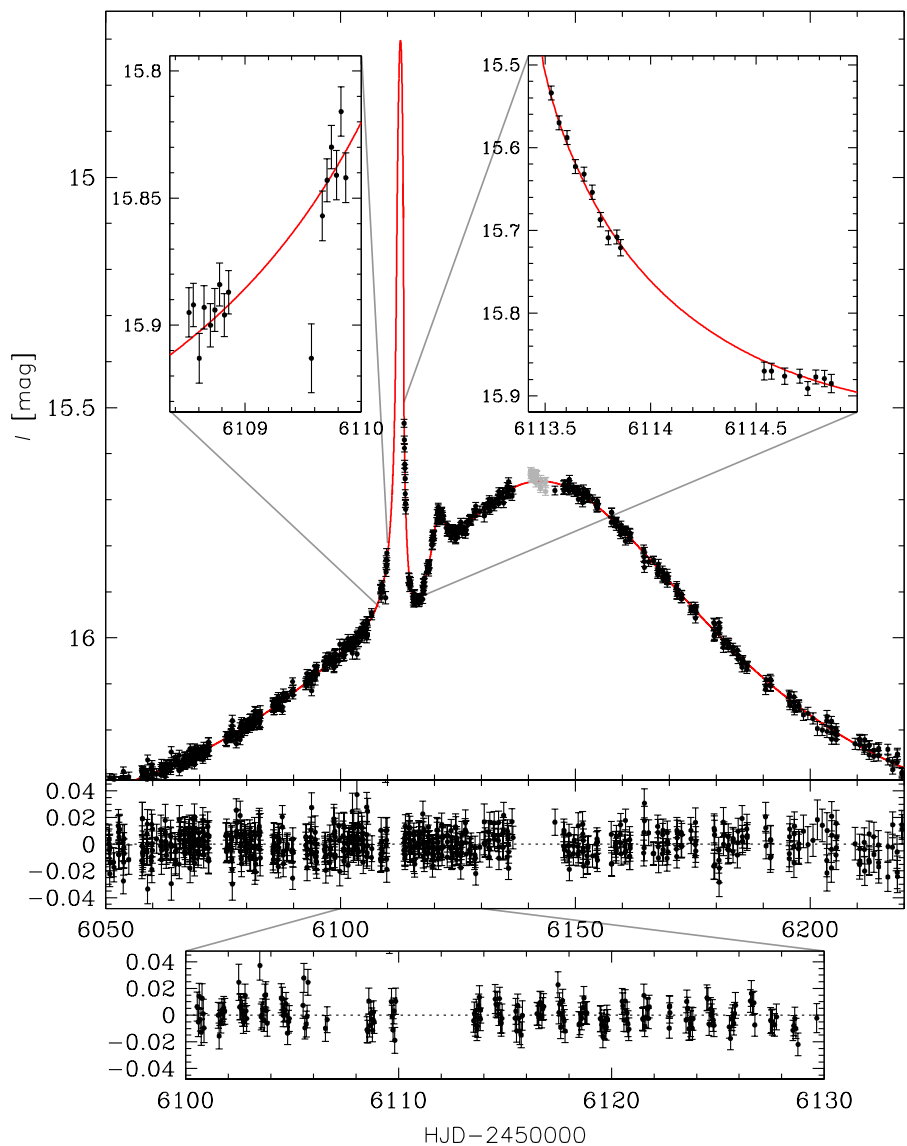}
\caption{Light curve of OGLE-2012-BLG-0406 planetary event.
The best fitted model is shown by red line. 
The gray points were rejected from the analysis because they were affected by Moon proximity. 
The insets show the reliability of the fit for the four nights closest to the brightest peak. 
The middle panel shows the residuals of the fit. 
The residuals for the time around anomalies are shown in the bottom panel.
\label{fig:lc}}
\end{figure}

The light curve of the event is presented in Figure~\ref{fig:lc}.
At the baseline $I = 16.459~{\rm mag}$ and $V = 19.268~{\rm mag}$. 
The uncertainties of photometry were scaled 
as is typically done in the analysis of microlensing events: 
$\sigma_{\rm new} = 1.5\sqrt{\sigma_{\rm old}^2 + (4~{\rm mmag})^2}$ \citep[see, e.g.,][]{skowron11}.
We transformed photometry to the standard system using \citet{szymanski11} photometric maps.

\section{Microlensing model} 

The light curve clearly shows anomalies and cannot be fitted with the standard point source-point lens Paczy\'nski light-curve ($\Delta\chi^2\approx10000$).
The rough estimates of the event properties can be found using analytical approximations presented by \citet{han06} (see Appendix). 
We start our search with a static binary-lens model and run it on a wide grid of $q$, $s_0$ and $\alpha_0$ parameters
(planet-host mass ratio and separation, and the source trajectory angle, respectively).\footnote{
Authors were aware of the microlensing model fit by Y. K. Jung et al., which was based on preliminary OGLE and follow-up data, but the presented analysis is independent and does not rely on the mentioned model.
}  
The limb-darkening coefficient $u_I = 0.450$ ($\Gamma_I = 0.353$) was assumed. 
We try to fit models in which the source passes the lensing star on the same side as planetary caustic 
as well as the opposite side. 
Only the former ones resulted in acceptable fits. 
After the first well fitting models were found the microlensing parallax ($\pi_{{\rm E},N}$, $\pi_{{\rm E},E}$) 
and lens orbital motion ($\gamma_\parallel$, $\gamma_\perp$) were taken into account. 
We check models with both positive and negative values of the impact parameter $u_0$ to examine a well-known degeneracy \citep{smith03}. 
See \citet{skowron11} for detailed description of standard microlensing parameters. 
The source was passing the caustic for much shorter time than the orbital period of the system, thus the lens orbital motion is approximated by two components of instantaneous velocity. 
The magnification is calculated using the hexadecapole approximation \citep{gould08,pejcha09} except the points with the highest magnification. 
These are four points with ${\rm HJD'}$ in the range $6113.56501 - 6113.68359$.
For them we use the ``mapmaking'' method \citep{dong06,dong09b}. 
We further enhanced the mapmaking procedure by calculating point source magnifications for all the images 
that are outside the map. 
This allows us to make the map smaller and for calculations to be more accurate. 
The map is placed around the planet because it caused the anomaly and has the radius of $\theta_{\rm E}q^{1/2}$. 
Mapmaking implicitly assumes $\gamma_\parallel = (ds/dt)/s_0 = 0$; 
thus, we constructed the map at ${\rm HJD'} = 6113.5846$. 
We have checked that ignoring $ds/dt$ during the $0.12~{\rm d}$ time interval affects the calculated magnifications at a level that is much smaller than the photometric errors. 

\begin{table}
\begin{center}
\caption{Best-fit model parameters.\newline \label{tab:res}}
\footnotesize{
\begin{tabular}{lcrrrrrr}
\tableline\tableline
\multicolumn{1}{c}{\tiny Quantity} &
\multicolumn{1}{c}{\tiny Unit} &
\multicolumn{1}{c}{\tiny Parallax and} &
\multicolumn{1}{c}{\tiny Parallax and} &
\multicolumn{1}{c}{\tiny Parallax and} &
\multicolumn{1}{c}{\tiny Orbital motion} &
\multicolumn{1}{c}{\tiny Parallax} &
\multicolumn{1}{c}{\tiny No Parallax} \\
 & & \multicolumn{1}{c}{\tiny orbital motion} & \multicolumn{1}{c}{\tiny orbital motion} & \multicolumn{1}{c}{\tiny orbital motion} & \multicolumn{1}{c}{\tiny only} & \multicolumn{1}{c}{\tiny only} & \multicolumn{1}{c}{\tiny and no} \\ 
 & & & \multicolumn{1}{c}{\tiny with priors} & ($u_0<0$) & & & \multicolumn{1}{c}{\tiny orbital motion}\\ 
\tableline
$\chi^2$ &    &  $1651.96$ &   &  $1665.56$ &  $1676.42$ &  $1701.95$ &  $1712.83$ \\
 ${\rm dof}$  &  &  $1652$  &  &  $1652$  &  $1650$  &  $1650$  &  $1648$   \\
$\frac{f_b}{f_s}$  &    &  $0.039$ &  $0.035$ &  $0.006$ &  $-0.028$ &  $-0.072$ &  $-0.069$ \\
 &  &  $\pm0.021$  &  $\pm0.019$  &  $\pm0.020$  &  $\pm0.020$  &  $\pm0.006$  &  $\pm0.006$   \\
$t_0-6141$  &  d   &  $0.548$ &  $0.565$ &  $0.583$ &  $0.676$ &  $0.736$ &  $0.593$ \\
 &  &  $\pm0.090$  &  $\pm0.076$  &  $\pm0.093$  &  $\pm0.075$  &  $\pm0.048$  &  $\pm0.031$   \\
$u_0$  &  $r_{\rm E}$   &  $0.5065$ &  $0.5072$ &  $-0.5169$ &  $0.5257$ &  $0.5441$ &  $0.5425$ \\
 &  &  $\pm0.0068$  &  $\pm0.0059$  &  $\pm0.0065$  &  $\pm0.0068$  &  $\pm0.0024$  &  $\pm0.0022$   \\
$t_{\rm E}$  &  d  &  $64.33$ &  $64.28$ &  $61.94$ &  $63.79$ &  $62.25$ &  $62.63$ \\
 &  &  $\pm0.49$  &  $\pm0.44$  &  $\pm0.81$  &  $\pm0.54$  &  $\pm0.27$  &  $\pm0.16$   \\
$\pi_{{\rm E}, N}$  &    &  $-0.118$ &  $-0.118$ &  $0.232$ &  -  &  $-0.026$ &  -  \\
 &  &  $\pm0.026$  &  $\pm0.026$  &  $\pm0.078$  &  &  $\pm0.021$  &   \\
$\pi_{{\rm E}, E}$  &    &  $0.046$ &  $0.046$ &  $0.020$ &  -  &  $0.029$ &  -  \\
 &  &  $\pm0.010$  &  $\pm0.010$  &  $\pm0.008$  &  &  $\pm0.010$  &   \\
$\rho$  &    &  $0.0109$ &  $0.0106$ &  $0.0093$ &  $0.0074$ &  $0.0097$ &  $0.0098$ \\
 &  &  $\pm0.0013$  &  $\pm0.0009$  &  $\pm0.0013$  &  $\pm0.0014$  &  $\pm0.0004$  &  $\pm0.0004$   \\
$q$  &  $10^{-3}$  &  $6.26$ &  $6.01$ &  $5.40$ &  $4.17$ &  $5.85$ &  $5.78$ \\
 &  &  $\pm0.78$  &  $\pm0.41$  &  $\pm0.64$  &  $\pm0.53$  &  $\pm0.10$  &  $\pm0.08$   \\
$s_0$  &  $r_{\rm E}$   &  $1.3134$ &  $1.3157$ &  $1.3289$ &  $1.3453$ &  $1.3514$ &  $1.3500$ \\
 &  &  $\pm0.0091$  &  $\pm0.0058$  &  $\pm0.0085$  &  $\pm0.0087$  &  $\pm0.0017$  &  $\pm0.0016$   \\
$\alpha_0$  &  deg    &  $47.98$ &  $48.01$ &  $-47.40$ &  $49.29$ &  $49.35$ &  $49.58$ \\
 &  &  $\pm0.29$  &  $\pm0.25$  &  $\pm0.60$  &  $\pm0.18$  &  $\pm0.14$  &  $\pm0.09$   \\
$\gamma_\parallel$  &  ${\rm yr}^{-1}$  &  $-0.46$ &  $-0.42$ &  $-0.22$ &  $0.06$ &  -  &  -  \\
 &  &  $\pm0.16$  &  $\pm0.09$  &  $\pm0.15$  &  $\pm0.15$  &  &   \\
$\gamma_\perp$  &  ${\rm rad}~{\rm yr}^{-1}$  &  $0.57$ &  $0.42$ &  $0.29$ &  $-0.90$ &  -  &  -  \\
 &  &  $\pm0.51$  &  $\pm0.27$  &  $\pm0.48$  &  $\pm0.44$  &  &   \\
\tableline
\end{tabular}
}
\end{center}
\end{table}

The parameters of the best-fitting models are presented in Table~\ref{tab:res}. 
The quantity $\rho$ is the source source size relative to Einstein ring radius 
and $f_b/f_s$ denotes blend to source flux ratio. 
The model with both the binary motion and the microlensing parallax gives the smallest $\chi^2$. 
It has $u_0>0$ and the solution with $u_0<0$ is worse by $\Delta\chi^2=13.6$. 
We report parameters of both models. 
The positive/negative $u_0$ ambiguity is not fully removed; thus, we present analysis of both cases. 
For the best-fitting model without lens orbital motion as well as the one without parallax, we present the $u_0>0$ solution only, consistently with the lowest $\chi^2$ model. 
In these cases, the positive/negative $u_0$ degeneracy is very severe (maximum $\Delta\chi^2=2.4$).
Significant negative blending \citep{park04} suggests that the last three models presented are not correct solutions, i.e., 
neglecting parallax or lens orbital motion results in wrong parameter estimates. 
The parameters were fitted using Monte Carlo Markov Chains.
We checked how inclusion of data collected during four nights around $t_0$ influences the parameters fitted. 
The 21 additional points increased $\chi^2$ by $25.1$. 
The largest change in values of parameters fitted was only $0.33\sigma$ in case of $t_{\rm E}$.

Table~\ref{tab:res} reveals that there is a degeneracy between $\pi_{{\rm E}, N}$ 
(which is a proxy for component of $\bpi_{{\rm E}}$ perpendicular to the direction of the projected Sun's acceleration) 
and $\gamma_{\perp}$. 
This degeneracy is well known \citep{skowron11,batista11}. 
There is also a suggestion that the analyzed event is a subject to degeneracy between $\pi_{{\rm E}, E}$ and $\gamma_{\parallel}$, which was presented by \citet{park13}. 
The reason of the second degeneracy is not well understood. 

\begin{figure}
\epsscale{.85}
\plotone{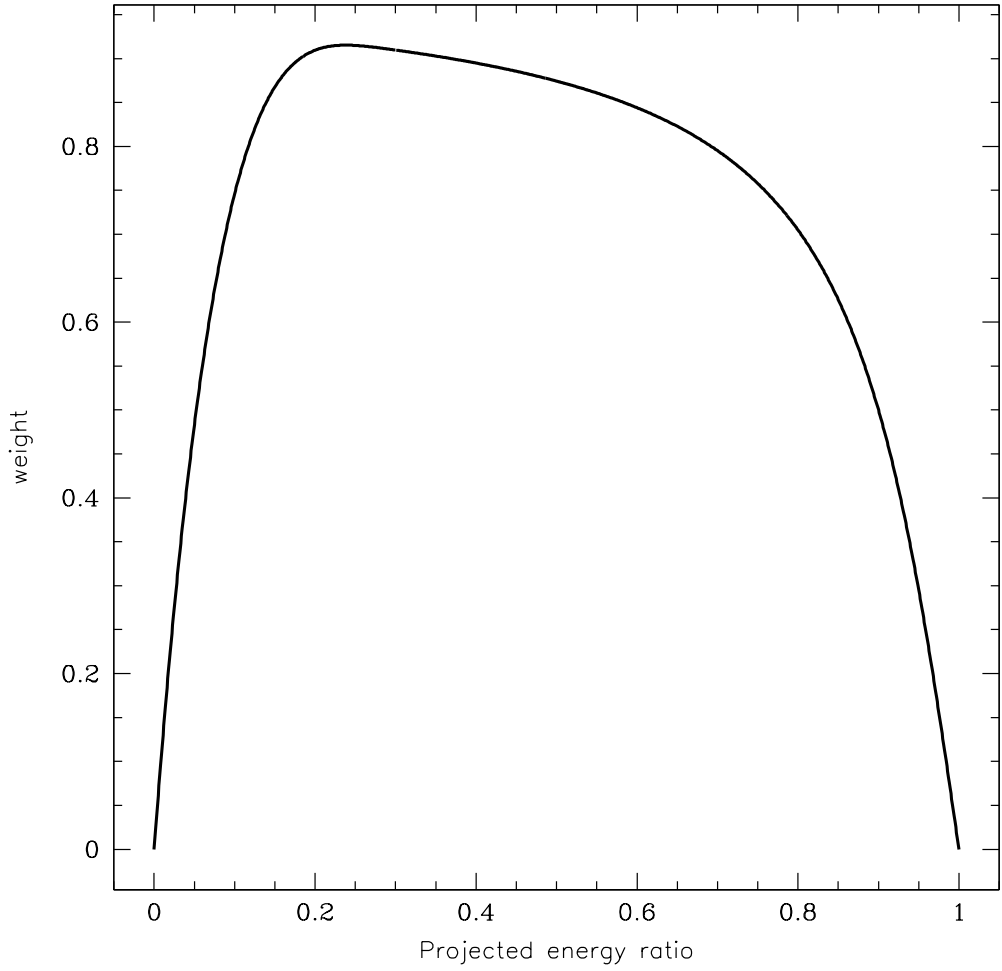}
\caption{Priors constraining the projected energy ratio ${E_{\perp,{\rm kin}}}/{E_{\perp,{\rm pot}}}$.
\label{fig:prior}}
\end{figure}

The model that includes both the microlensing parallax and orbital motion of the lens gives $\chi^2$ smaller by $24.5$ than any other model. 
This shows that the effects of the lens orbital motion are significant yet not
strong enough to warrant a full Keplerian orbital analysis. In order
to use information that comes with a detection of an orbital motion,
we put arbitrary priors on the ratio between projected kinetic and potential
energy:
\begin{equation}
\frac{E_{\perp,{\rm kin}}}{E_{\perp,{\rm pot}}} = \frac{(\bgamma s_0 D_l\theta_{\rm E})^2}{{2GM}/{r_{\perp}}},
\end{equation}
where $D_l$ is lens distance, $\theta_{\rm E}$ is the angular Einstein radius (see next Section),
and $r_{\perp}$ is projected star planet separation.
The priors\footnote{
The priors are defined as 
$f({E_{\perp,{\rm kin}}}/{E_{\perp,{\rm pot}}} = x)=\arctan\left(10-10x\right)/12$ for $x\geq0.4$
For $x\leq0.25$, we assumed $f(x)=2\beta x$ and $\beta=0.2$.
For $0.25<x<0.4$, the third order polynomial is used that at $0.25$ and $0.4$ has values and first derivatives equal to values of corresponding functions defined above.
} 
are presented in Fig.~\ref{fig:prior}. 
The ratio ${E_{\perp,{\rm kin}}}/{E_{\perp,{\rm pot}}}$ must be smaller than unity so that the solution is bound. 
Values close to unity are obviously disfavored. 
We also disfavored small values of ${E_{\perp,{\rm kin}}}/{E_{\perp,{\rm pot}}}$ because they would correspond to the planet lying close to the lens-observer line, which should be small part of the planetary orbit. 
We note that for a flat distribution in logarithm of semimajor axis (\"Opik's Law) and circular orbits the distribution of projected energy ratio is $2\beta{E_{\perp,{\rm kin}}}/{E_{\perp,{\rm pot}}}$, where 
$\beta$ is the exponent of semimajor axis distribution. 
This defines our prior for small values of ${E_{\perp,{\rm kin}}}/{E_{\perp,{\rm pot}}}$.
Comparison of the third and fourth columns of Table~\ref{tab:res} 
reveals that parameter estimates are consistent and 
the priors did not significantly affect the precision of most of the parameters. 
The exceptions are the mass ratio and parameters of the lens orbital motion. 
Analysis of additional data for this system should allow one to determine whether such priors provide more accurate parameters values. 

\begin{figure}
\epsscale{.764}
\plotone{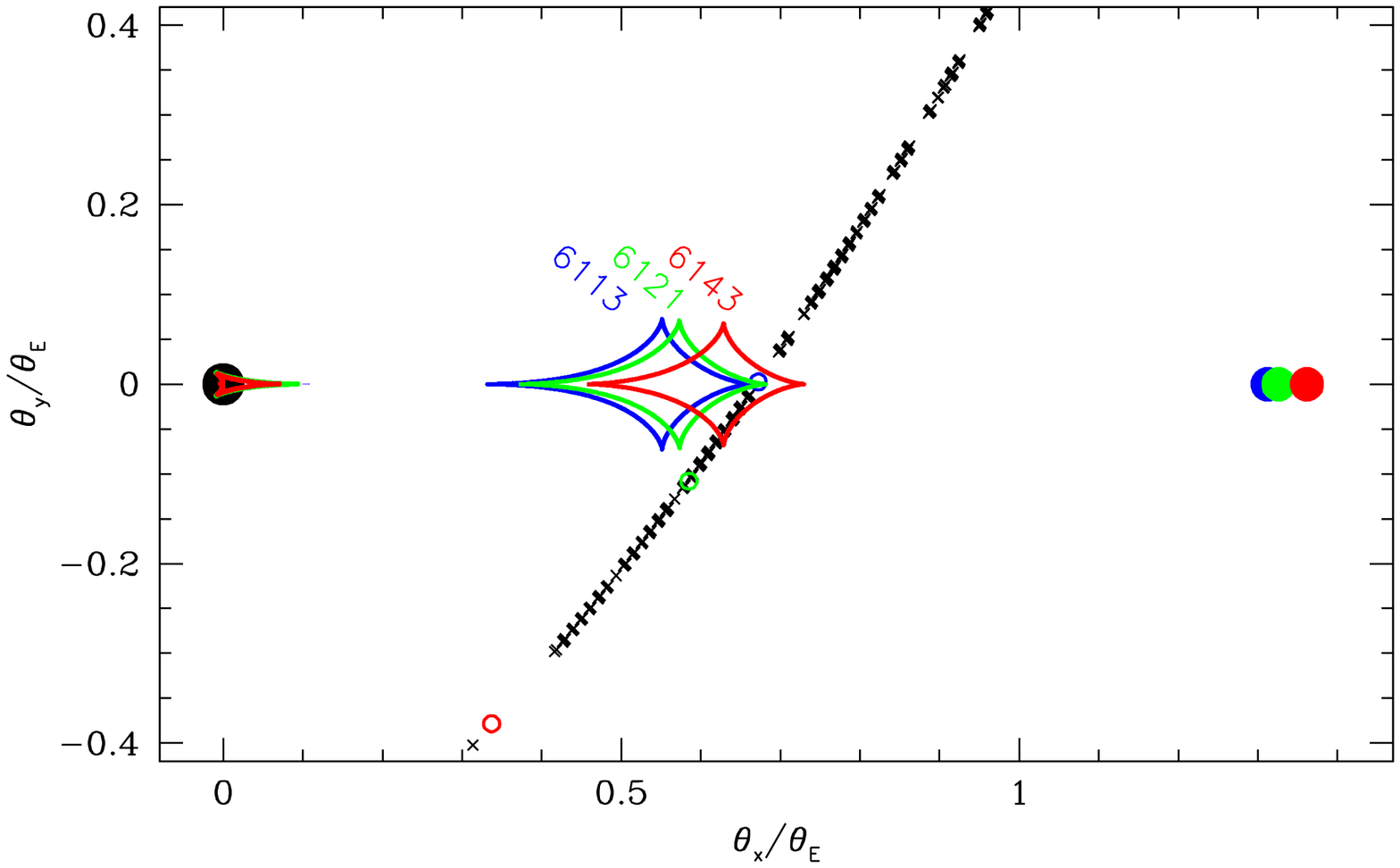}
\caption{Trajectory of the source. 
Crosses represent the positions of the source when the images were taken. 
Caustics, planet position (filled circle), and source position (empty circle) are plotted for three epochs
(truncated HJD is provided): 
the closest approach to one cusp (blue), the closest approach to the other cusp (green),  
and the closest approach to the lensing star (red).
The filled black circle represents the lensing star.
\label{fig:traj}}
\end{figure}

The trajectory of the source relative to the caustics and lensing system is presented in Figure~\ref{fig:traj}. 
Because of the nonzero $\gamma_\perp$, the caustics and the planet are plotted for three selected moments in time. 

\begin{figure}
\epsscale{.74}
\plotone{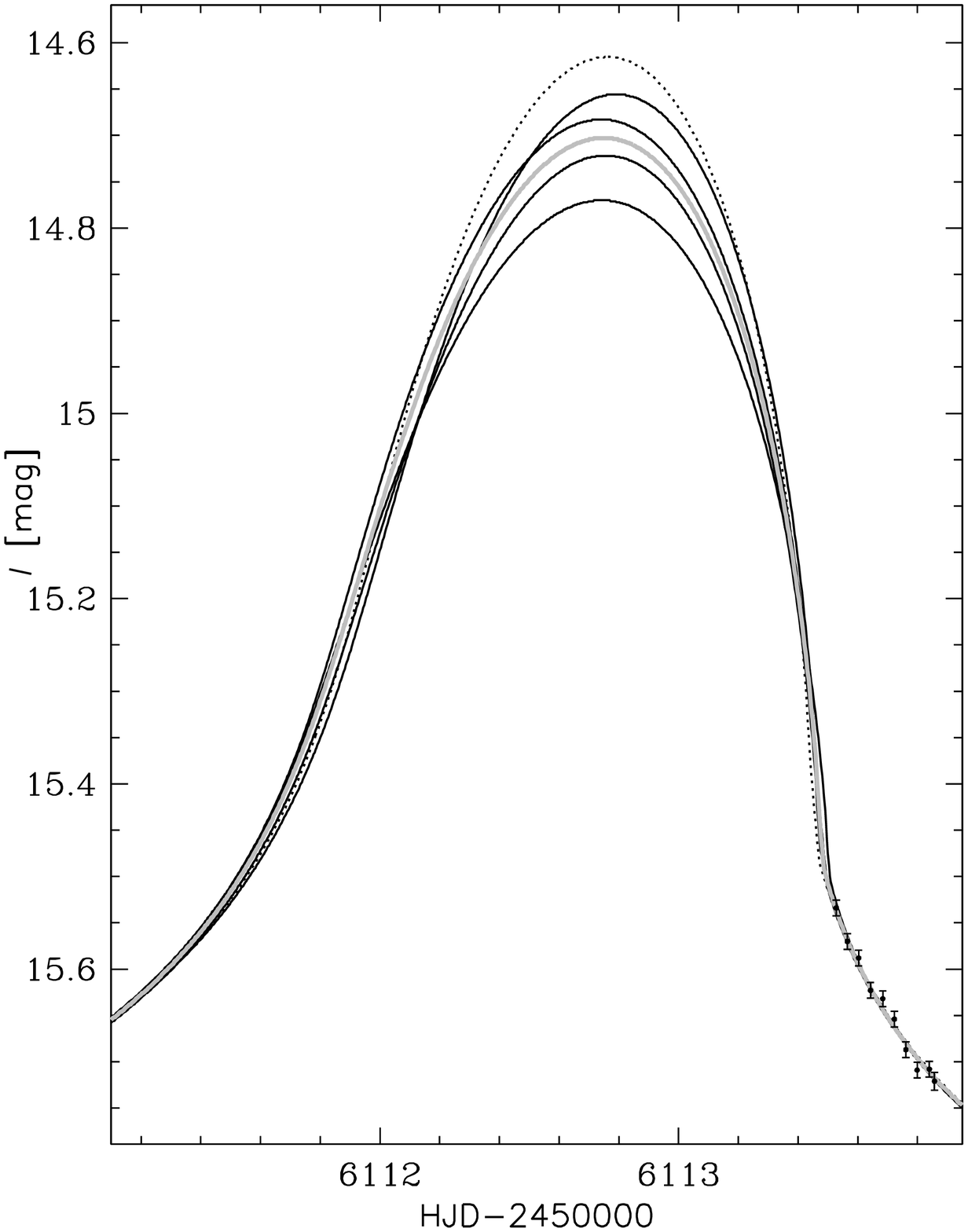}
\caption{Predicted brightness variations during the first cusp crossing. 
Gray line presents the best fitting model with $u_0>0$ and 
four black lines show randomly chosen models with $\Delta\chi^2 = 9$.  
Dotted curve illustrates the best fitting model with $u_0<0$. 
In order to present which part of the predicted light curve is well constrained 
we present the first observing night after the peak using points. 
The last measurement before the peak was taken at ${\rm HJD'} = 6109.87$.
\label{fig:pred}}
\end{figure}

In Figure~\ref{fig:pred}, we present the predicted brightness variations 
for the best-fitting models with $u_0>0$ (gray thick line) and $u_0<0$ (dotted line). 
We also present four randomly chosen models with $\Delta\chi^2 = 9$ to illustrate the uncertainties in our model. 

\section{Physical properties} 

\begin{figure}
\epsscale{.85}
\plotone{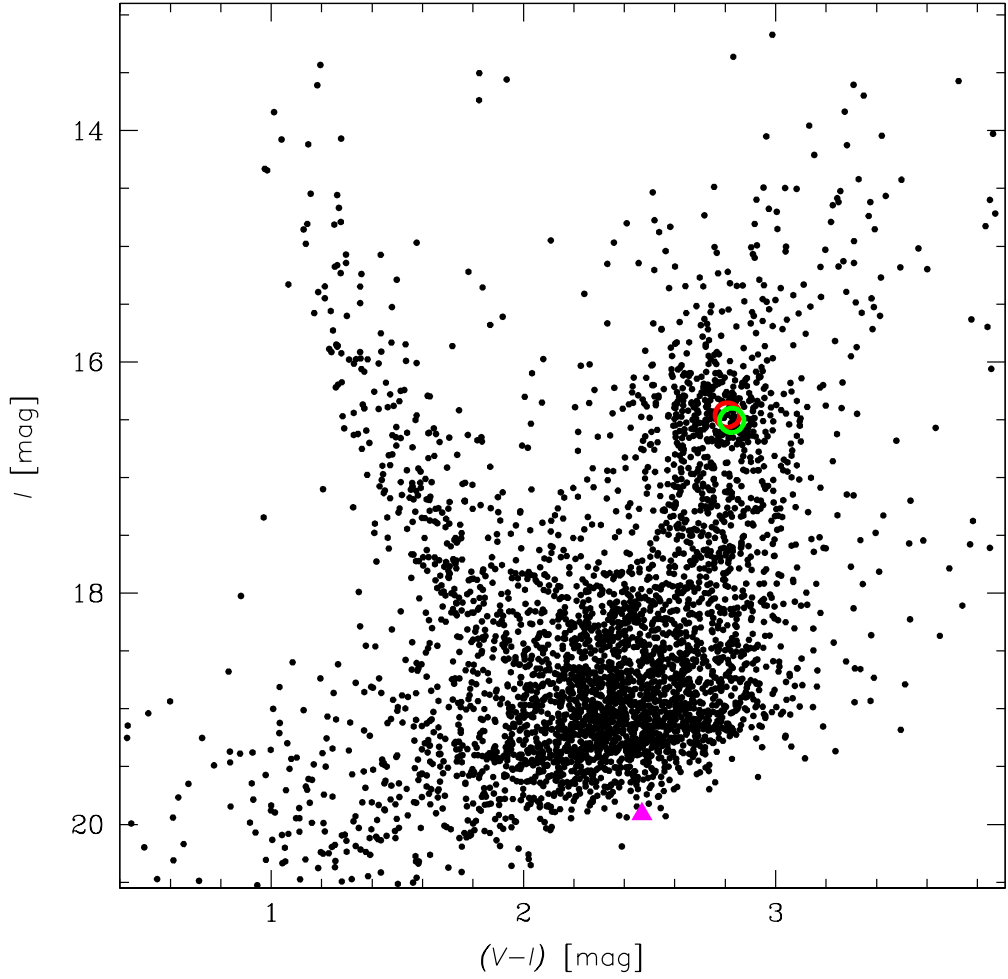}
\caption{Color-magnitude diagram for stars within $1.5\arcm$ of the event. 
The green circle shows the position of the source star as derived from the microlensing model, 
while red circle presents the  
total light observed at the baseline. 
Magenta triangle presents position of the blend. 
\label{fig:CMD}}
\end{figure}

In order to constrain the physical properties of the system one must first estimate 
the angular Einstein radius $\theta_{\rm E}$. 
We do this by standard microlensing technique \citep{yoo04b}. 
The position of the source star and the blended object on the color-magnitude diagram is presented in Figure~\ref{fig:CMD}. 
The position of the source star is consistent with bulge red clump.
This figure also presents the brightness and color of the blended light. 
There is one star of $I$-band brightness between $19.5$ and $20.5~{\rm mag}$ every $3~{\rm arcsec^2}$ in the Galactic bulge \citep{holtzman98}.
Because of high stellar density and large uncertainty in blending flux,  
we cannot unambiguously attribute the blending flux to the lens.

\begin{table}
\begin{center}
\caption{Derived physical parameters. The $u_0>0$ solution is favored (Tab.~\ref{tab:res}).\newline \label{tab:phys}}
\footnotesize{
\begin{tabular}{lcrr}
\tableline\tableline
\multicolumn{1}{c}{Parameter} &
\multicolumn{1}{c}{Unit} &
\multicolumn{1}{c}{$u_0>0$ solution} &
\multicolumn{1}{c}{$u_0<0$ solution} \\
\tableline
$\theta_{\rm E}$ & mas   & $0.57\pm0.07$ & $0.68\pm0.09$\\
$M$ & $M_{\odot}$        & $0.59\pm0.17$ & $0.48\pm0.20$\\
$M_p$ & $M_{\rm Jup}$   & $3.8\pm1.2$   & $2.6\pm1.2$\\
$D_L$ & kpc              & $5.1\pm0.4$   & $3.6\pm1.3$\\
$\mu_{\rm geo}$ & mas/yr & $3.2\pm0.4$   & $3.9\pm0.6$\\
$r_{\perp}$ & AU         & $3.8\pm0.6$   & $3.2\pm1.2$\\ 
${E_{\perp,{\rm kin}}}/{E_{\perp,{\rm pot}}}$ &
                         & $0.67$        & $0.15$\\ 

\tableline
\end{tabular}
}
\end{center}
\end{table}

The intrinsic color ($(V-I)_{\rm RC,0} = 1.06~{\rm mag}$) and brightness ($I_{\rm RC,0} = 14.47~{\rm mag}$) of the red clump were taken from \citet{bensby11} and \citet{nataf13b}, respectively. 
By combining these with the measured color and brightness of the red clump in the vicinity of the event, we obtained the reddening and extinction of $E(V-I) = 1.70~{\rm mag}$ and $A_I = 1.96~{\rm mag}$. 
The dereddened color and brightness of the source star were measured using microlensing model: $(V-I)_0 = 1.12~{\rm mag}$ and $I_0 = 15.66~{\rm mag}$. 
Using the relations by \citet{bessell88}, we obtain the source color $(V-K)_0 = 2.60~{\rm mag}$, 
which correlates well with the source size \citep{kervella04giant}. 
The measured angular size of the source is $\theta_* = 6.3 \pm 0.3 {\rm \mu as}$. 
The source size relative to Einstein ring radius $\rho = \theta_* / \theta_{\rm E} = 0.0109 \pm 0.0012$ 
from $u_0>0$ model implies $\theta_{\rm E} = \theta_*/\rho = 0.57\pm0.07~{\rm mas}$. 

The value of $\theta_{\rm E}$ allows estimating parameters of the lens in physical units. 
As there remained little ambiguity between $u_0>0$ and $u_0<0$ models, we present the physical parameters of the lens for both solutions in Table~\ref{tab:phys}.
We assume the source parallax $\pi_s = 0.125~{\rm mas}$ 
since the source is a bulge red clump star (Figure~\ref{fig:CMD}). 
The geocentric proper motion of the lens is $\mu_{\rm geo} = \theta_{\rm E}/t_{\rm E} = 3.3 \pm 0.4~{\rm mas/yr}$. 
The lens proper motion is consistent with the estimated distance. 
The proper motion corrected to the heliocentric frame is $\bmu_{\rm helio} = \left[3.2\pm1.0,1.6\pm0.4\right]~{\rm mas/yr}$ (North, East). 
If the next class of adaptive optics systems are build, they should be able to resolve the lens and the source in ten years.

\begin{figure}
\plotone{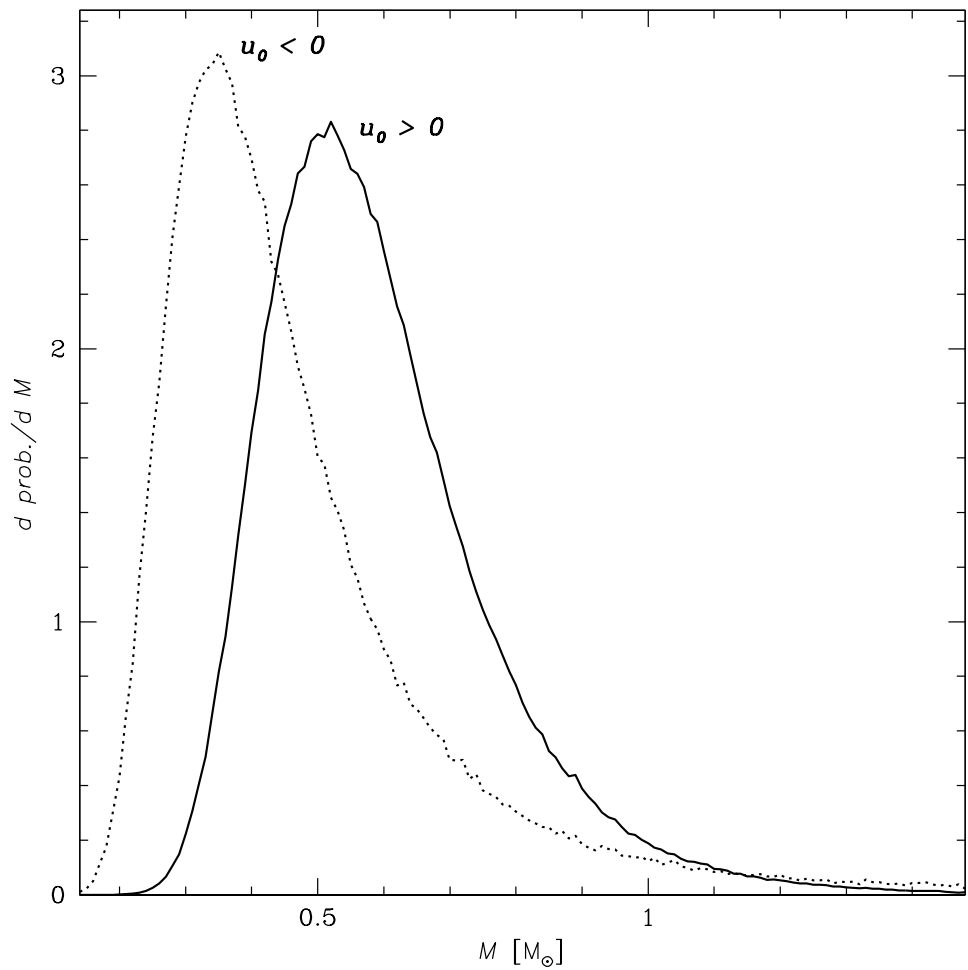}
\caption{Probability distribution of the lens mass for two best-fitting models.
\label{fig:mass}}
\end{figure}

The mass of the lens is \citep{gould00b}: 
\begin{equation}
M = \frac{\theta_{\rm E}}{\kappa\pi_{\rm E}};~\kappa=\frac{4G}{c^2{\rm AU}}\approx8.1~\frac{\rm mas}{M_\odot}.
\end{equation}
The probability distribution of the host mass is shown in Fig.~\ref{fig:mass} for both $u_0>0$ and $u_0<0$ solutions. 
The boundary mass of M/K dwarfs of $\approx0.6~M_{\odot}$ is close to the peak of distribution for $u_0>0$ solution. 
We conclude that the star is either K or M dwarf. 
The long tail of the mass distribution is caused by the fact that $\pi_{\rm E}$ is detected at only $5\sigma$ and $3\sigma$ for $u_0>0$ and $u_0<0$, respectively. 
The priors on the projected energy ratio did not significantly affect the lens mass estimate. 
The values of kinetic to potential projected energy ratio \citep{batista11} 
are consistent with a bound system.

The values of all the quantities presented in this section together with $M_p$, $D_L$ and $r_{\perp}$ are given in Table~\ref{tab:phys}. 
Since the  positive/negative $u_0$ degeneracy is not fully removed, we present results for both solutions.

\section{Discussion and summary} 

We presented the microlensing detection of a super-Jupiter orbiting low-mass star. 
The lens mass can be used to estimate how far from the star the snow line lies.
It is approximately $2.7~{\rm AU}(M/M_{\odot}) = 1.6~{\rm AU}$, which is significantly 
smaller than the derived projected separation of 
$3.8 \pm 0.6~{\rm AU}$.  
We note that OGLE-2012-BLG-0406Lb is the fourth 
super-Jupiter orbiting a low-mass star beyond the snow line discovered by the microlensing method. 
The existence of such objects is a challenge for the core accretion theory of planetary formation
and to our knowledge has not been addressed.
For a recent review on the core accretion theory see \citet{zhou12}.

Single telescope data were used to discover OGLE-2012-BLG-0406Lb, what makes this planet unique among microlensing discoveries with well characterized host lensing and planetary anomaly. 
We note that if the unfortunate Moon proximity to the event has not occurred during the cusp crossing, most probably the parameters of the lensing system would be derived more accurately. 
All the fitted model parameters can be verified by incorporating data from other ground-based telescopes and the EPOXI spacecraft, which also observed the event.

The presence of the planet caused two well-separated anomalies, each lasting a few days. 
One may ask why more such detections were not claimed before as microlensing surveys had a cadence of one day a decade ago. 
It should be noted that even though such a cadence should be enough to detect anomalies caused by similar planets in other events, it is not enough to give a high level of confidence. 
Only recently microlensing surveys achieved cadences of $15-60$ minutes 
over many square degrees in the Galactic bulge, 
which allows reliable detections of such planets. 
Microlensing planetary discoveries benefit most from the follow-up observations only if the central or resonant caustic is causing the anomaly \citep{griest98} 
as was the case for OGLE-2005-BLG-071 \citep{dong09a} and MOA-2011-BLG-387 \citep{batista11}. 
In the event analyzed here, the planetary caustic is causing the anomaly in the light curve. 
In such events, the anomaly is well separated from the lensing star, thus the magnification of the event before, and after the caustic approach is rather low and the approach may be long before or after the closest approach to the lensing star.
Follow-up groups typically do not densely monitor such events \citep{gould10}. 
Unlike planetary 
events with central-caustics perturbation,
planetary-caustic events do not suffer from ambiguity in planet-star
separation due to the discrete close-wide degeneracy. 
Thus, the
planet sample discovered from planetary caustics may provide cleaner
statistics in orbital separation distribution than those found in
high-magnification events. 
The OGLE-IV survey discovers around 2000 microlensing
events every year, and independent from the follow-up observations, it
alone will potentially provide a powerful probe on the distribution of
super-Jupiters orbiting low-mass stars beyond the snow line. 
The evidence for the existence of 
such planets 
is growing and puts a severe challenge for the core accretion theory of planetary formation.

\acknowledgments
We thank the anonymous referee for fruitful suggestions. 
The OGLE project has received funding from the European Research Council under the European Community’s Seventh Framework Programme (FP7/2007-2013)/ERC grant agreement No.~246678 to AU.
SD was supported through a Ralph E.~and Doris M.~Hansmann Membership at the IAS and NSF grant AST-0807444. 
SK was also supported by the Polish Ministry of Science and Higher Education (MNiSW) through the program ``Iuventus Plus'' award No. IP2011 026771. 
AG acknowledges supported by NSF grant AST 1103471. 

\appendix
\section{Microlensing model parameters from analytical approximation} 

\begin{figure}
\plotone{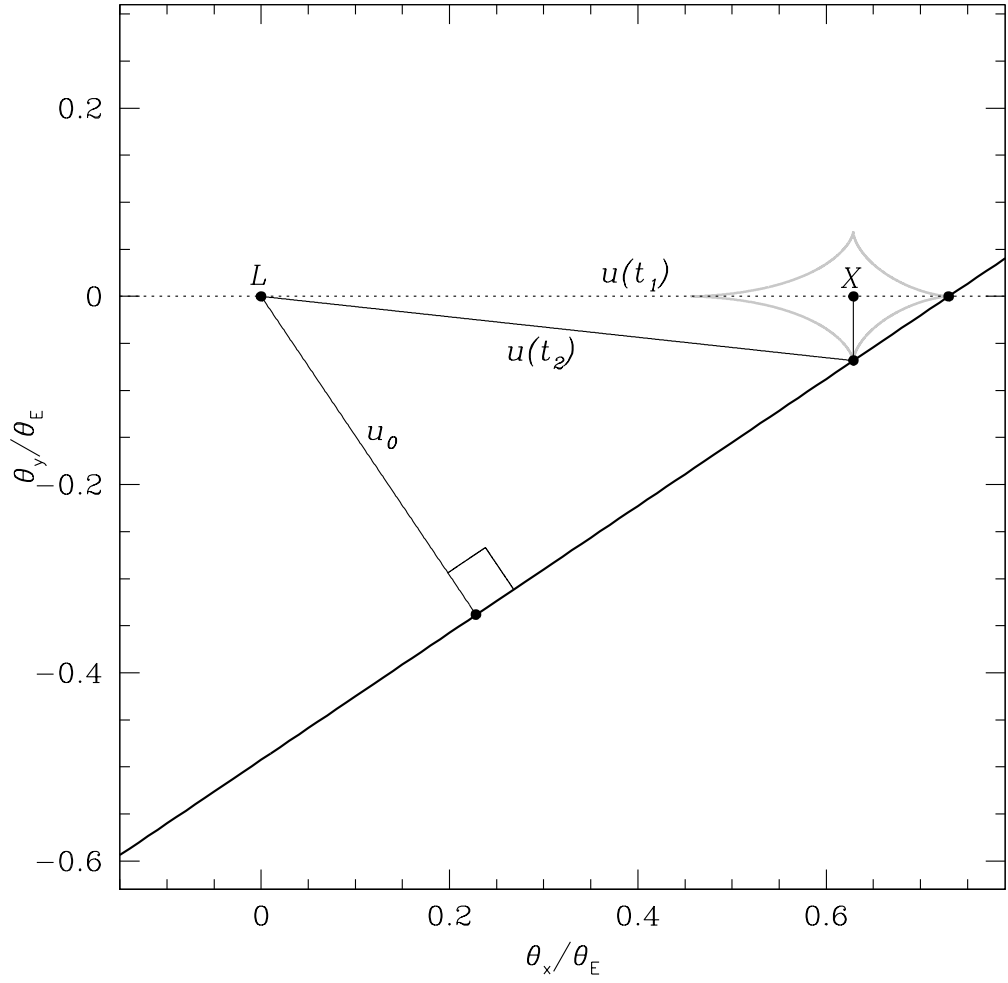}
\caption{Simplified geometry of the event. 
The gray line marks the planetary caustic. 
The dotted line is the binary axis. 
Source trajectory is presented by the thick line.
\label{fig:geo}}
\end{figure}

The idea that the observed characteristics of the planetary microlensing event give estimates of lens parameters 
 was presented long time ago \citep{gould92,gaudi97}. 
Here we present an example of such calculations, which yield 
basic point-lens ($u_0$, $t_{\rm E}$) and
static binary-lens parameters ($\alpha_0$, $s_0$, $q$). 
The rectilinear motion of the source and absence of blending are assumed.
The geometry of the source trajectory compared to the planetary caustic and the lens is presented in Figure~\ref{fig:geo}.

The brightness difference between the baseline and the time when the source is closest to the lens is $0.78~{\rm mag}$. 
Thus, the maximum magnification $A_{\rm max} = 2.05$. 
It is related to the minimum impact parameter $u_0$ via \citet{paczynski86} formula:
\begin{equation}
A_{\rm max} = \frac{u^2_0+2}{u_0\sqrt{u^2_0+4}}
\end{equation}
Inverting this relation yields $u_0 = 0.539$. 
The above equation can be used to estimate the magnification, 
and thus the brightness, of the event at epochs $t_0-t_{\rm E}$ and $t_0+t_{\rm E}$ 
where $t_0 = 6143.4$. 
The result is $16.09~{\rm mag}$. 
We can find the epochs in Figure~\ref{fig:lc} when this brightness was observed. 
Half of the time interval between these two epochs gives $t_{\rm E} = 62.5~{\rm d}$.
We note that the values of $u_0$ and $t_{\rm E}$ critically depend on the assumption that the blending can be neglected. 

Figure~\ref{fig:lc} can be used to find the epoch when the source approached 
the first ($t_1 = 6112.4~{\rm d}$) 
and the second cusp ($t_2 = 6120.9~{\rm d}$). 
This allows one to determine the distance of the source from the lens at these two epochs in the units of $\theta_{\rm E}$:
\begin{equation}
u(t) = \sqrt{\left(\frac{t-t_0}{t_{\rm E}}\right)^2+u_0^2}.
\end{equation}
It results in $u(t_1) = 0.73$ and $u(t_2) = 0.65$. 
This in turn can be used to find the value of 
$\alpha = \arcsin(u_0/u(t_1))= 47.4^{\circ}$.
However, small changes in the value of $\alpha$ can result in significant changes in the light curve shape and this value should be treated with caution. 

The light curve of OGLE-2012-BLG-0406 showed two anomalies, and we assume that each of them corresponded to the source passing exactly through the cusp. 
Figure~\ref{fig:traj} can be used to validate this assumptions in the fitted model. 
During the second anomaly, the distance between the source and the cusp was significant, but we are not able to make better analytical approximation near cusps \citep{pejcha09}. 

The value of $s_0$ can be found by calculating the distance between the lens and the position of the second cusp projected onto the binary axis (point $X$ in Figure~\ref{fig:geo}). 
Basic geometry gives a value of $0.61$. 
It is equal to $s_0 - 1/s_0$ thus $s_0 = 1.35$. 
The distance between the projected position of the second cusp and the first cusp is $2\sqrt{q}/\left(s\sqrt{s^2-1}\right)$ \citep{han06} and in the analyzed case equals $0.126$. 
Thus the mass ratio is 
$0.0059$.

The values found using this analytical approximation are close to the results presented in Table~\ref{tab:res} especially when compared to the model without microlensing parallax and orbital motion. 
In the case of events perturbed by the cusps of planetary caustics, such calculations should be useful for starting values in model fitting.

\bibliographystyle{apj}

\end{document}